\documentclass[]{spie}
\usepackage{makeidx}  
\usepackage{url}    
\usepackage{graphicx}

\def\BibPath{/Users/markus/Work/Tex/}
\title{Classification of Interstitial Lung Disease Patterns with Topological Texture Features}
\author{Markus B. Huber\supit{1}, 
		Mahesh Nagarajan\supit{1},
		Gerda Leinsinger\supit{2},
		Lawrence A. Ray\supit{3} and \\
              	Axel Wism\"uller\supit{1} 
\skiplinehalf
\supit{1}Departments of Imaging Sciences and Biomedical Engineering, University of Rochester, \\
		New York, United States; \\
\supit{2}Department of Radiology, University of Munich, Germany;\\
\supit{3}Carestream Health, Inc., Rochester, New York, United States
}
\authorinfo{Further author information: (Send correspondence to M.B.H.)\\
M.B.H.: E-mail: mbh@bme.rochester.edu, Telephone: +1-585-276-4776\\}
\begin{document}
%
%
\maketitle 
\begin{abstract}
Topological texture features were compared in their ability to classify morphological patterns known as 'honeycombing' that are considered indicative for the presence of fibrotic interstitial lung diseases in high-resolution computed tomography (HRCT) images. 
For 14 patients with known occurrence of honey-combing, a stack of 70 axial, lung kernel reconstructed images were acquired from HRCT chest exams.
A set of 241 regions of interest of both healthy and pathological (89) lung tissue were identified by an experienced radiologist. 
Texture features were extracted using six properties calculated from gray-level co-occurrence matrices (GLCM), Minkowski Dimensions (MDs), and three Minkowski Functionals (MFs, e.g. MF.euler). 
A $k$-nearest-neighbor ($k$-NN) classifier and a Multilayer Radial Basis Functions Network (RBFN) were optimized in a 10-fold cross-validation for each texture vector, and the classification accuracy was calculated on independent test sets as a quantitative measure of automated tissue characterization.
A Wilcoxon signed-rank test was used to compare two accuracy distributions and the significance thresholds were adjusted for multiple comparisons by the Bonferroni correction.
The best classification results were obtained by the MF features, which performed significantly better than all the standard GLCM and MD features ($p<0.005$) for both classifiers.
The highest accuracy was found for MF.euler (97.5\%, 96.6\%; for the $k$-NN and RBFN classifier, respectively).
The best standard texture features were the GLCM features 'homogeneity' (91.8\%, 87.2\%) and 'absolute value' 
(90.2\%, 88.5\%).
The results indicate that advanced topological texture features can provide superior classification performance in computer-assisted diagnosis of interstitial lung diseases when compared to standard texture analysis methods.
\end{abstract}
\keywords{texture analysis, quantitative image analysis, lung high resolution CT, supervised learning.}
%
%
\section{Introduction} 
%
%
The rapid progress in computed tomography (CT) acquisitions of the chest have been followed by a sharp increase
in research on computer analysis of thoracic CT scans and the development of computer-aided diagnosis systems~\cite{sluimer2006}.
In high-resolution CT (HRCT) images of the chest, the classification of pathological patterns of interstitial lung diseases (ILDs) is a complicated classification task which requires substantial expertise.
Such ILD disorders are characterized by specific  abnormal findings with a distinct texture-like appearance.
One morphological pattern related to the heterogenous group of ILDs is  'honeycombing'~\cite{uppaluri1999,uchiyama2003,boehm2008}.

HRCT images can provide an accurate assessment of lung tissue patterns that allow automated analysis tools to provide quick information to the radiologists (or even non-specialists).
A reliable computer-aided diagnosis could improve the radiologists' efficiency and avoid surgical lung biopsies for some patients.
Furthermore, the radiologist's ability to interpret HRCT depends on their expertise and might be subjective, whereas an automated classification of lung patterns is consistent and reproducible. 

A computer-aided diagnosis system should be used in a pre-screening step in order to label suspicious images and 
regions that include pathological patterns.
This setting incorporates the valuable experience and judgement of the radiologist, improves the radiologist's productivity and reduces reading fatigue.
One approach to design such an image-based, automated detection tool is to extract relevant texture features from regions of interest (ROI) and the use of supervised machine learning techniques to classify the ROI as healthy or pathological pattern.
The focus of this paper is the question whether recently developed texture features provide a better classification of lung patterns.
 
Standard measures like features extracted from gray-level occurrence matrices (GLCM)~\cite{haralick1973} calculate second-order statistics of the gray-level distributions.
Another popular approach to texture analysis is to assess global fractal dimensions, e.g. the Minkowski Dimension\cite{maragos1994}).
Both methods have been widely used in the medical image analysis problem, e.g. in the analysis of radiographs~\cite{jiang1999a,huber2009}
Recently developed texture features like Minkowski Functionals (MF)~\cite{hadwiger1957,mecke1991, michielsen2001} provide information about topology of the underlying gray-level pattern within the ROI  and include higher-order statistics.
Rather than representing the ROIs with statistical properties or fractal dimensions, the MF feature vectors represent the 
shape and the connectivity of the gray levels in the lung pattern. 
This concept has been successfully applied previously in the to assessment bone quality with various imaging modalities~\cite{boehm2006,raeth2006,huber2009}.

The purpose of this work is to compare a set of standard texture features with this new approach to texture analysis and to train two classifiers in order to detect healthy and pathological ROIs in lung HRCT images.


\section{Data} 

\begin{figure}
  \begin{center}
	 \includegraphics[height=5cm] {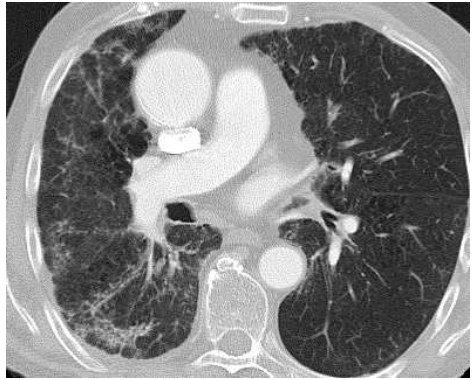}    
    \caption{High-resolution computed tomography (HRCT) image of the lung). 
       }
    \label{FIG:lung}
  \end{center}
\end{figure}

\begin{figure}
  \begin{center}
      \begin{tabular}{cc}
    \includegraphics[width=4cm] {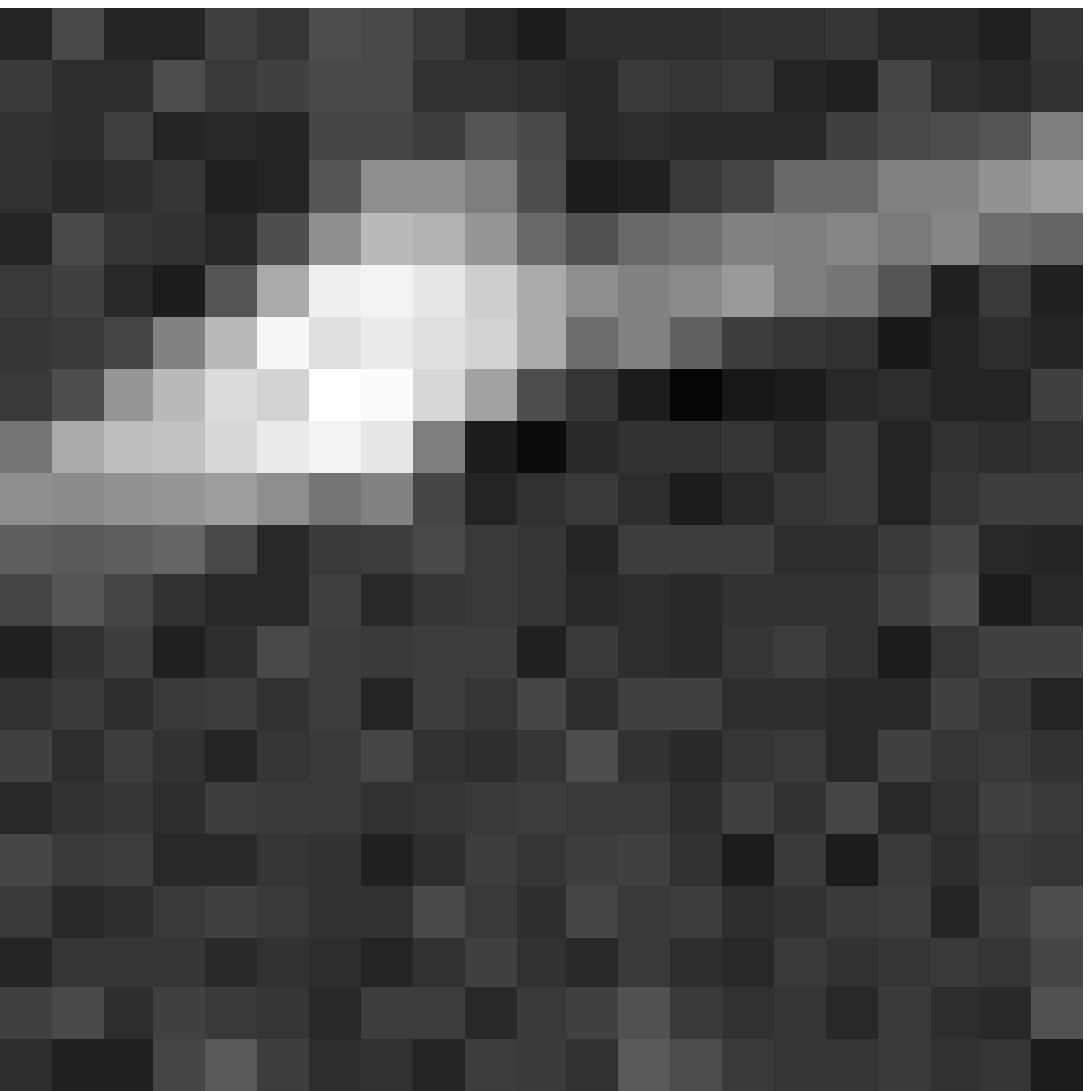} &
    \includegraphics[width=4cm] {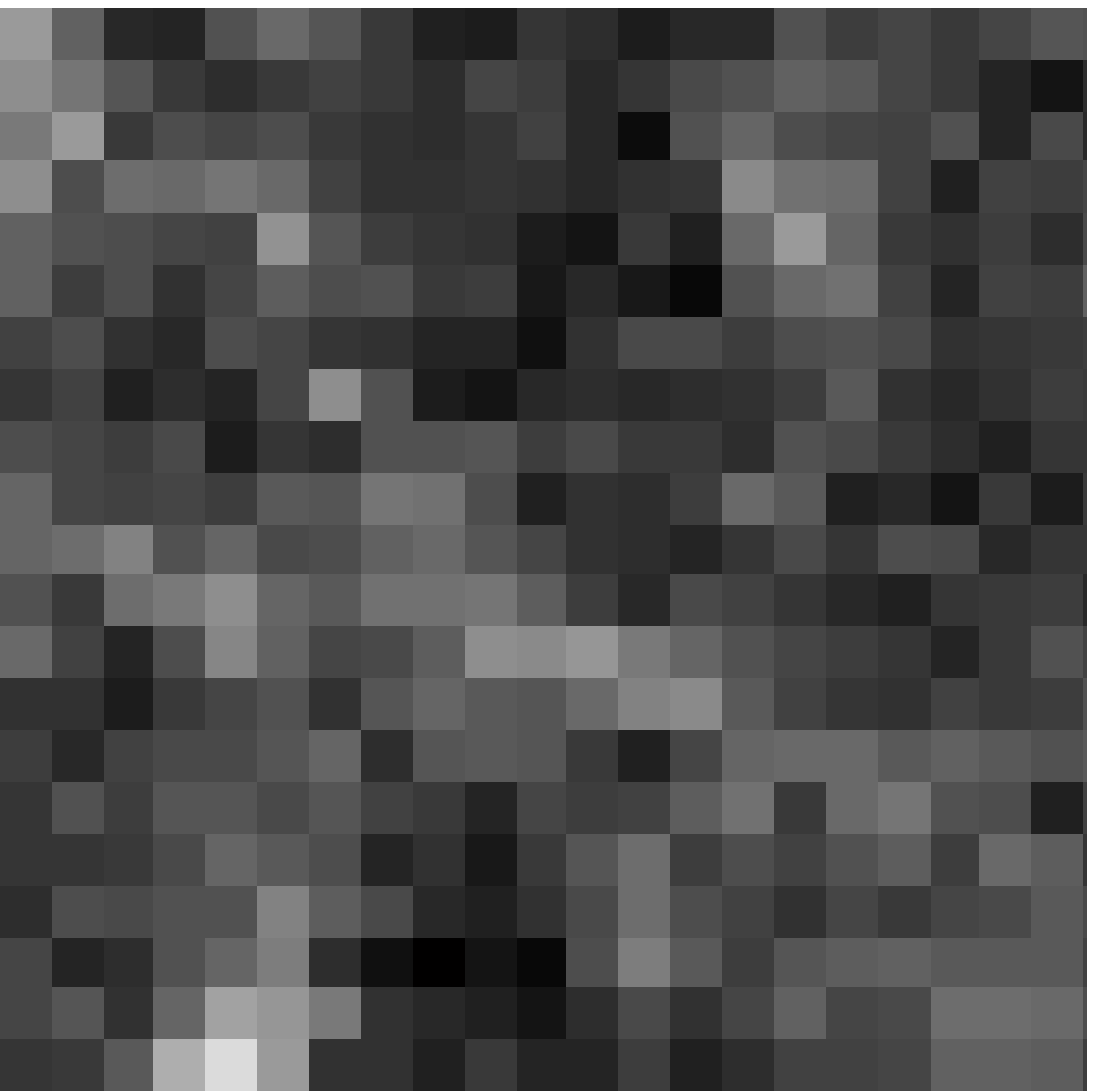}
    \end{tabular}    
    \caption{ROIs extracted from HRCT images with healthy lung tissue (left) and pathological lung tissue (right). }
    \label{FIG:rois}
  \end{center}
  \end{figure}
The used data set consists of chest scans of 14 patients with known occurrence of honeycombing patterns.
For each patient, a stack of 70 axial images reconstructed with a lung kernel from high-resolution computed tomography 
(HRCT) of the chest was assessed by an experienced radiologist (Fig.~\ref{FIG:lung}). 
The resulting images are comparable to classical HRCT with the exception that reconstructions
of a small number of slice positions are also available.
All images have a slice thickness of 1mm, an in-plane pixel size of 0.69x0.69 mm, 
and a slice distance of 5 mm. 
In this data set, 241 square, 2-dimensional ROIs with an edge length of 21 pixels were defined
in 152 healthy and 89 pathological pattern occurrences in lung tissue (Fig.~\ref{FIG:rois}).
In each of these ROIs, the set of  the following feature vectors of texture descriptors  were
calculated.

\section{Methods}
\subsection{Texture Features}

Gray level co-occurrence matrices (GLCM)~\cite{haralick1973,haralick1979} are second order histograms that estimate the joint probability of a gray level pixel $g_i$ and gray level pixel $g_j$ for a given pixel spacing and direction. 
From the GLCMs, the least correlated and those most frequently used statistical features were calculated~\cite{anys1995} (see Appendix):
absolute value (absval), entropy, contrast, energy, correlation (correl), and homogeneity (homogen).
In each ROI, the gray levels were rescaled (min-max-scaling) to 256 gray level values and texture features were calculated for $n_{\rm dir}=4$ offset directions $\theta=\{0^{\circ}, 45^{\circ}, 90^{\circ}, 135^{\circ}\}$ and five pixel spacings $s_p=\{1,2,3,4,5\}$. Pixels pairs were excluded from the GLCM if the neighbor pixel was outside the ROI. 
	
	Fractal properties were assessed with the global Minkowski Dimension (MD) \cite{maragos1994,jiang1999a,jiang1999b}, a robust estimator of fractal dimensions based on morphological operations. If a two-dimensional image $I$ is regarded as a gray-level surface, a ÒvolumeÓ difference $V_g(s)$ can be calculated between a dilated and eroded conversion of the the image $I$, i.e. $(I \oplus sg)$ and $(I \ominus sg)$ respectively, for a given structuring element $g$ and a given scale $s$:
\begin{equation}
V_g(s)=\sum_{m=0}^M \sum_{n=0}^N [(I \oplus sg) - (I \ominus sg)].
\end{equation}	
	
The global MD can then be defined as~\cite{maragos1994}
	
\begin{equation}{\label{eq:MD_def}}
{\rm MD}=\lim_{s\rightarrow\infty}\frac{\log[V_g(s)/s^3]}{\log(1/s)}.
\end{equation}
	
In practice, the MD features were assessed by plotting the numerator and the denominator in Equation (\ref{eq:MD_def}) and
finding the slope using a least-square fit.	
For the calculation, 6 structuring elements were used: 2 isotropic ( a square and a disk) and 4 anisotropic elements pointing in four different directions ($0^{\circ}$, $45^{\circ}$, $90^{\circ}$, $135^{\circ}$).
	
Morphological image analysis characterizes binary images in terms of shape (geometry) and connectivity (topology) given by the Minkowski Functionals (MFs) \cite{hadwiger1957,mecke1991, michielsen2001, boehm2008,huber2009}. 
It can be shown that those morphological properties of the objects in the binary image are completely described by the corresponging MFs~\cite{hadwiger1957,mecke1991}. 
By defining a set of thresholds, the gray-level images of the lung pattern can be represented as a set of binary images.
For each of those 2D binary images, three MF features can defined: area, perimeter (boundary length) and the Euler characteristic $\chi$ (connectivity):
\begin{eqnarray}
{\rm MF.area}  &=& n_s \\
{\rm MF.perim} &=& -4n_s+2n_e \\
{\rm MF.euler} &=& n_s-n_e+n_v
\end{eqnarray}
where $n_s$ is the total number of white pixels, $n_e$ the number of edges, and $n_v$ the number of vertices.
These three MFs are functions of the gray level threshold, which characterize the underlying topological features of the image texture. 
The gray level distributions in each ROI was divided into 100 quantiles which were then used as thresholds to calculate the three MF features.
The resulting three feature vectors for each ROI was used as input for the classification task.
 
\subsection{Classification}

After the the calculation of the texture properties, the machine learning task consists of the classification  between healthy and pathological lung patterns in a supervised learning step.
Two classifiers were used in this study, a $k$-Nearest Neighbor ($k$-NN), which reflects learning by density estimation, and
Multilayer Radial Basis Function Neural Network (RBFN), which is a nonlinear numerical approach for this classification task~\cite{duda2000}.
In order to identify unknown ROIs, models are created from known labeled data in the training phase. 
The goal is to find the optimal classifier parameters that model the best boundaries between the two lung classes for each texture feature.
For the $k$-NN classifier, the free parameter was the number of nearest neighbors $k$, for the RBFN classifier, the shape parameter of the Gaussian radial basis function.
The performance of optimization was measured by the accuracy, i.e. the total number of correctly classified data points, as a quantitative measure of automated tissue characterization.

In order to generalize the resulting performance, the model parameters were calculated with a training set of labeled ROIs and the accuracy was calculated on an independent test set of ROIs. 
In one iteration, the data set is divided into a training set (70\%) and a test set (30\%). 
The training set is used to determine the best classifier parameter in a 10-fold cross-validation. 
This classifier model is then applied on the independent test set to calculate the accuracy of one iteration. 
This procedure was repeated 50 times resulting in an accuracy distribution for each texture feature and classifier. 
A Wilcoxon signed-rank test was used to compare two accuracy distributions and the significance thresholds were
adjusted for multiple comparisons by the Bonferroni correction.
 
The RBF implementation was taken from the Neural Network Toolbox v6.0.1 in Matlab R2008b (TheMathWorks, Natick, MA); the texture analysis, the optimization and the statistical analysis were implemented in Matlab as well.

\section{Results}\label{sec:results}

\begin{table}[tb]
\begin{center}
\begin{tabular}[t]{lccccc c ccccc}
\hline\\ [-1ex]
Features     & \multicolumn{5}{c}{ $k$-NN}             && \multicolumn{5}{c}{ RBFN}  \\
		& Accu. [\%] && Sens.  [\%] && Spec.  [\%] && Accu.  [\%] && Sens.  [\%] && Spec.  [\%]\\ [1ex]
\hline \hline\\[-2ex]

GLCM.absval   &$ 90.2 \pm 3.4 $&&$ 90.1 \pm 5.8 $&&$ 90.1 \pm 5.5 $& &$ 88.5 \pm 3.7 $&&$ 96.0 \pm 2.9 $&&$ 75.9 \pm 7.6 $ \\
GLCM.entropy   &$ 85.2 \pm 3.6 $&&$ 85.3 \pm 7.6 $&&$ 85.1 \pm 4.8 $& &$ 71.1 \pm 4.8 $&&$ 69.6 \pm 6.9 $&&$ 73.5 \pm 8.7 $ \\ 
GLCM.contrast  &$ 82.8 \pm 4.4 $&&$ 84.5 \pm 6.9 $&&$ 82.0 \pm 6.4 $& &$ 81.0 \pm 5.0 $&&$ 87.9 \pm 6.8 $&&$ 70.2 \pm 9.3 $ \\[0.5ex]
GLCM.correl    &$ 77.4 \pm 5.2 $&&$ 77.0 \pm 9.1 $&&$ 78.0 \pm 7.6 $& &$ 80.5 \pm 5.0 $&&$ 84.2 \pm 7.0 $&&$ 73.9 \pm 9.7 $ \\
GLCM.energy    &$ 86.2 \pm 4.0 $&&$ 85.2 \pm 6.6 $&&$ 86.8 \pm 4.8 $& &$ 83.9 \pm 3.8 $&&$ 83.0 \pm 5.6 $&&$ 86.0 \pm 7.2 $ \\
GLCM.homogen   &$ 91.8 \pm 2.9 $&&$ 88.3 \pm 5.6 $&&$ 94.0 \pm 3.5 $& &$ 87.2 \pm 4.3 $&&$ 87.5 \pm 5.6 $&&$ 86.6 \pm 8.0 $ \\[0.5ex]
MD             &$ 86.7 \pm 4.2 $&&$ 84.2 \pm 7.7 $&&$ 88.4 \pm 4.8 $& &$ 84.3 \pm 3.2 $&&$ 84.7 \pm 5.3 $&&$ 83.6 \pm 7.8 $ \\[0.5ex]
MF.area        &$ 96.6 \pm 2.0 $&&$ 96.9 \pm 3.4 $&&$ 96.5 \pm 2.5 $& &$ 93.3 \pm 2.7 $&&$ 93.4 \pm 3.7 $&&$ 93.4 \pm 4.5 $ \\
MF.perim       &$ 96.1 \pm 2.2 $&&$ 94.8 \pm 4.2 $&&$ 97.0 \pm 2.4 $& &$ 95.3 \pm 1.9 $&&$ 96.2 \pm 2.7 $&&$ 93.9 \pm 4.1 $ \\
MF.euler       &$ 97.5 \pm 1.4 $&&$ 97.5 \pm 3.0 $&&$ 97.5 \pm 1.8 $& &$ 96.6 \pm 2.5 $&&$ 96.5 \pm 3.2 $&&$ 96.8 \pm 3.6 $ \\[0.5ex]

\hline\\
\end{tabular}
\caption{Accuracy, sensitivity, and specificity for each texture feature vector and each classifier. The means and the standard deviations
of the 50 iterations are reported.}\label{TAB:Results}
\end{center}
\end{table}

\begin{figure}
  \begin{center}
      \includegraphics[width=12cm] {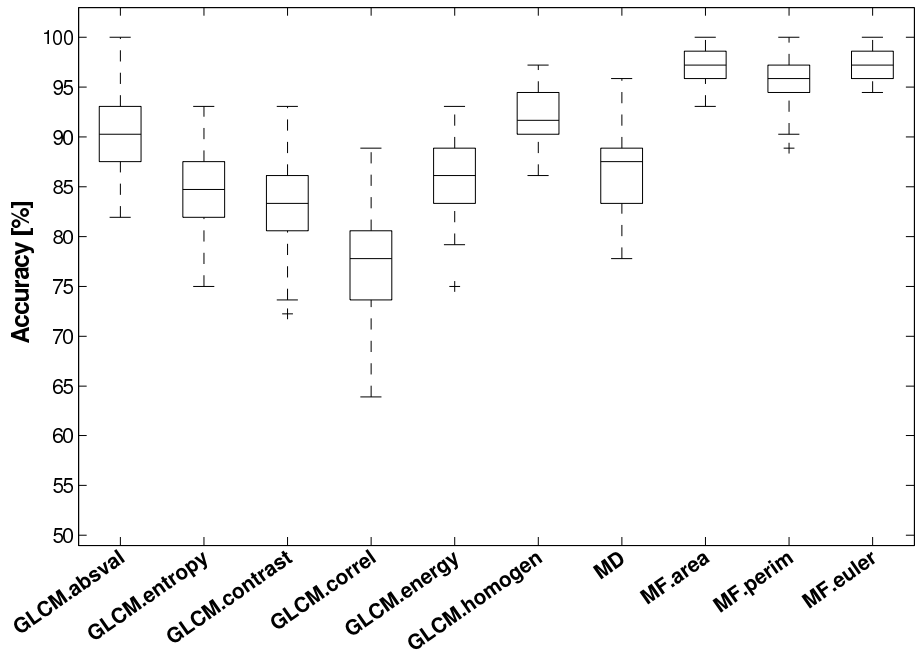} \\[8ex]
       \includegraphics[width=12cm] {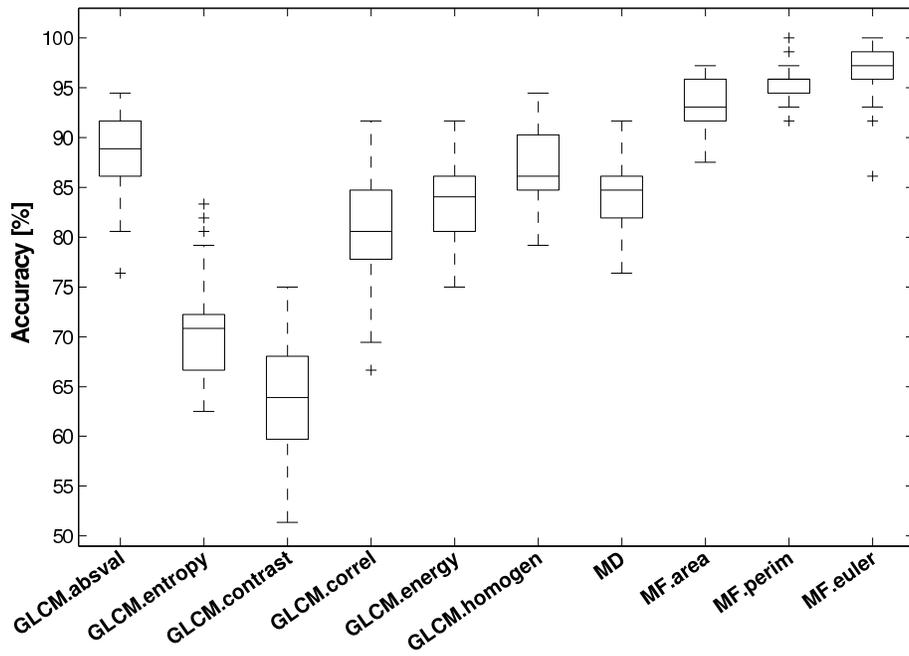} 
    \caption{Accuracies of the texture features using the $k$-NN (top) and the RBFN (bottom) classifier. 
    In each box, the central mark corresponds with the median, the edges of the box are the 25th and 75th percentile, and the error bars extend to the most extreme data points not considering outliers which are plotted individually (crosses).
       }
    \label{FIG:boxplot_knn}
  \end{center}
\end{figure}

In Tab.~\ref{TAB:Results}, the measured accuracy, sensitivity, and specificity for each texture feature vector and both classifiers are presented.
The best classification results were obtained by the MF features, which performed significantly better than the
GLCM and MD features ($p<0.001$) for both classifiers (Fig.~\ref{FIG:boxplot_knn}).
The highest accuracy of 97.5\% and 96.6\% was found for the MF.euler feature for the $k$-NN and RBFN classifier,
respectively. The best standard texture features were GLCM.absval and GLCM.homogen, which had  significantly higher accuracies than the other GLCM and MD features ($p<0.005$ for $k$-NN, $p<0.001$ for RBFN).

The superior classification performance of the MF features is also reflected in the corresponding high specificity and sensitivity values (Tab.~\ref{TAB:Results}).
Only GLCM.absval had a comparable sensitivity of 96.0\% but a has small specificity (RBFN), and only GLCM.homogen had a high specificity of 94\% but a small sensitivity ($k$-NN).

\section{Discussion}\label{sec:discussion}

In this study, recently developed texture features were used to classify a pathological pattern in lung HRCT images.
The MF features extract topological information from the image patterns
and had a significantly higher classification accuracy than established GLCM and MD texture features. 
The set up of this study allows an easy implementation of a clinical setting and the classification
can reach a high accuracy. 

Instead of extracting one feature from the set of MF~\cite{boehm2008},
feature vectors were created that represent the complete topological of 
the MF as function of the threshold.
An additional new approach was to use those high-dimensional feature vectors to train the $k$-NN and the RBFN
classifiers in order to classify the lung patterns.
This combination of topological texture analysis and pattern recognition techniques produced
high classification accuracy that can be used in computer-aided diagnosis applications.
Because the patches in this study were small, the analysis could be also used to segment
the complete lung tissue.
Each lung pixel can be classified based on the local surrounding tissue  and the results
of this classification can be visualized to the radiologists by marking pathological areas.
The machine learning step of the presented analysis may even be improved by comparing
more classifiers
as different supervised learning techniques could also affect the performance of the texture features in the classification
task~\cite{depeursinge2008}.

In addition, a different binning method was applied to calculate the MF features by  using
quantiles instead of bins of constant width~\cite{boehm2008} to sample the 
MF properties as a function of the threshold.
This approach guarantees that each bin is represented by a minimal
number of gray levels and empty vector entries are avoided. 
This binning provides a more consistent and statistically stable representation of the 
MF features especially for small ROIs like used in this work, 

The superior classification performance  of the MF features correspond with the recent results
on chest HRCT studies~\cite{boehm2008}.
In current research, MF features have also been successfully used to assess bone quality in magnetic resonance images
\cite{boehm2003}, simulations~\cite{raeth2006}, multidetector CT scans~\cite{boehm2006}, 
and radiographs~\cite{huber2009}.
The approach of the MF features to represent the shape and the connectivity of the lung pattern in
a quantifiable manner seems to be well-suited for the pathological patterns produced by ILDs disorders.

An advantage of this method is that it can be easily implemented and applied in a clinical setting 
where it can provide high classification performances in detecting those pathological patterns.
It is our hope that this will improve the radiologists' ability to accurately classify ILDs.
However, to complete the validation of clinical applications, further investigation is necessary.
For completely automated differential diagnosis of ILDs, more information will be required to be 
incorporated into the computer system.
In this study, a defined set of acquisition and reconstruction parameters was used to assess the results. 
In future studies, the influence of the these parameters on the accuracy of the texture features
should be investigated.
Due to the fact that classification is thin-sclice CT scans is a difficult task even for experts, more radiologists
should be involved to define a reference standard for the machine learning step.

Furthermore, a limiting factor of this study might be that only one typical pathological 
lung pattern, i.e. honeycombing, was classified.
This represents only one of many interstitial
lung disease patterns that are common in clinical chest CT. 
In addition, a final objective is not only to classify between healthy and diseased tissue, but to 
differentiate between various pathological lung patterns~\cite{depeursinge2008}.
The influence of those different pathological patterns on the accuracy of the 
texture features  will be investigated in future studies.  
A further improvement of this approach is also the analysis of three dimensional volumes
instead of two dimensional areas.
\section{Conclusion}
In conclusion, the presented results indicate that advanced topological texture features can provide 
superior classification performance in computer-assisted diagnosis of interstitial
lung diseases compared to standard texture methods. 
However, future studies are necessary to validate these results in a clinical setting and to provide a reliable
differential diagnosis.

\section{Appendix}\label{sec:appendix}

For a gray-level occurrence matrix $P$ at a given pixel distance $s_p$ and offset angle $\theta$, 
\begin{equation}
P(i,j)=P(i,j,s_p,\theta),
\end{equation}
the following definitions for the GLCM features were used:
\begin{eqnarray*}
\mbox{Absolute Value}  &=& \sum_{i=1}^N \sum_{j=1}^N |i-j| P(i,j) \\
\mbox{Entropy} &=&  \sum_{i=1}^N \sum_{j=1}^N  P(i,j) \log_2 P(i,j) \\
\mbox{Contrast}  &=& \sum_{i=1}^N \sum_{j=1}^N (i-j)^2 P(i,j) \\
\mbox{Energy}  &=& \sum_{i=1}^N \sum_{j=1}^N  [P(i,j)]^2\\
\mbox{Correlation}  &=& \sum_{i=1}^N \sum_{j=1}^N [(i-\mu_x)(j-\mu_y) P(i,j)]/(\sigma_x \sigma_y) \\
\mbox{Homogeneity}  &=& \sum_{i=1}^N \sum_{j=1}^N P(i,j)/(1+|i-j|)
\end{eqnarray*}
where $\mu_x$ and $\sigma_x$ are the mean and the standard deviation of the rows in the GLCM, respectively, and
$\mu_y$ and $\sigma_y$  are the corresponding values for the columns.

\bibliography{\BibPath/091117_BibList}

\begin{thebibliography}{10}

\bibitem{sluimer2006}
Sluimer, I., Schilham, A., Prokop, M., and van Ginneken, B., ``Computer
  analysis of computed tomography scans of the lung: a survey,'' {\em Medical
  Imaging, IEEE Transactions on}~{\bf 25},  385--405 (April 2006).

\bibitem{uppaluri1999}
Uppaluri, R., Hoffman, E.~A., Sonka, M., Hartley, P.~G., Hunninghake, G.~W.,
  and McLennen, G., ``{Computer Recognition of Regional Lung Disease
  Patterns},'' {\em Am. J. Respir. Crit. Care Med.}~{\bf 160}(2),  648--654
  (1999).

\bibitem{uchiyama2003}
Uchiyama, Y., Katsuragawa, S., Abe, H., Shiraishi, J., Li, F., Li, Q., Zhang,
  C.-T., Suzuki, K., and Doi, K., ``Quantitative computerized analysis of
  diffuse lung disease in high-resolution computed tomography,'' {\em Medical
  Physics}~{\bf 30}(9),  2440--2454 (2003).

\bibitem{boehm2008}
Boehm, H.~F., Fink, C., Attenberger, U., Becker, C., Behr, J., and Reiser, M.,
  ``Automated classification of normal and pathologic pulmonary tissue by
  topological texture features extracted from multi-detector ct in 3d,'' {\em
  Eur Radiol}~{\bf 18}(12),  2745--55 (2008).

\bibitem{haralick1973}
Haralick, R.~M., Shanmugam, K., and Dinstein, I., ``Textural features for image
  classification.,'' {\em IEEE Trans. Syst., Man. Cybern.}~{\bf 3}(6),
  610--621 (1973).

\bibitem{maragos1994}
Maragos, P., ``Fractal signal analysis using mathematical morphology,'' {\em
  Adv. Electron. Electron. Phys.}~{\bf 88},  199--246 (1994).

\bibitem{jiang1999a}
Jiang, C., Giger, M.~L., Chinander, M.~R., Martell, J.~M., Kwak, S., and Favus,
  M.~J., ``Characterization of bone quality using computer-extracted
  radiographic features,'' {\em Med. Phys.}~{\bf 26}(6),  872--879 (1999).

\bibitem{huber2009}
Huber, M.~B., Carballido-Gamio, J., Fritscher, K., Schubert, R., Haenni, M.,
  Hengg, C., Majumdar, S., and Link, T.~M., ``Development and testing of
  texture discriminators for the analysis of trabecular bone in proximal femur
  radiographs,'' {\em Medical Physics}~{\bf 36}(11),  5089--5098 (2009).

\bibitem{hadwiger1957}
Hadwiger, H.,  [{\em Vorlesungen Ÿber Inhalt, OberflŠche und Isoperimetrie (in
  German)}{\nolinebreak\hspace{0.1em}]}, Springer, Berlin (1957).

\bibitem{mecke1991}
Mecke, K.~R. and Wagner, H., ``Euler characteristic and related measures for
  random geometric sets,'' {\em Journal of Statistical Physics}~{\bf 64}
  (1991).

\bibitem{michielsen2001}
Michielsen, K. and Raedt, H.~D., ``Integral-geometry morphological image
  analysis,'' {\em Physics Reports}~{\bf 347}(6),  461 -- 538 (2001).

\bibitem{boehm2006}
Boehm, H.~F., Link, T.~M., Monetti, R.~A., Kuhn, V., Eckstein, F., Raeth,
  C.~W., and Reiser, M., ``Analysis of the topological properties of the
  proximal femur on a regional scale: evaluation of multi-detector ct-scans for
  the assessment of biomechanical strength using local minkowski functionals in
  3d,'' {\em Medical Imaging 2006: Image Processing}~{\bf 6144}(1),  61446X,
  SPIE (2006).

\bibitem{raeth2006}
Raeth, C.~W., Mueller, D., Link, T.~M., Boehm, H., and Monetti, R.,
  ``Quantifying changes in the bone microarchitecture using
  minkowski-functionals and scaling vectors: a comparative study,'' {\em
  Medical Imaging 2006: Image Processing}~{\bf 6144}(1),  61440C, SPIE (2006).

\bibitem{haralick1979}
Haralick, R.~M., ``Statistical and structural approaches to texture,'' {\em
  Proc. IEEE}~{\bf 67}(5),  786--804 (1979).

\bibitem{anys1995}
Anys, H. and He, D.-C., ``Evaluation of textural and multipolarization radar
  features for crop classification.,'' {\em IEEE Transactions on Geoscience and
  Remote Sensing}~{\bf 33}(5),  1170--1181 (1995).

\bibitem{jiang1999b}
Jiang, C., Pitt, R.~E., Bertram, J.~E., and Aneshansley, D.~J., ``Fractal-based
  image texture analysis of trabecular bone architecture,'' {\em Med Biol Eng
  Comput}~{\bf 37}(4),  413--8 (1999).

\bibitem{duda2000}
Duda, R.~O., Hart, P.~E., and Stork, D.~G.,  [{\em Pattern
  Classification}{\nolinebreak\hspace{0.1em}]}, Wiley-Interscience Publication
  (2000).

\bibitem{depeursinge2008}
Depeursinge, A., Iavindrasana, J., Hidki, A., Cohen, G., Geissbuhler, A.,
  Platon, A., Poletti, P.~A., and Muller, H., ``Comparative performance
  analysis of state-of-the-art classification algorithms applied to lung tissue
  categorization,'' {\em J Digit Imaging}  (2008).

\bibitem{boehm2003}
Boehm, H.~F., Raeth, C., Monetti, R.~A., Mueller, D., Newitt, D., Majumdar, S.,
  Rummeny, E., Morfill, G., and Link, T.~M., ``Local 3d scaling properties for
  the analysis of trabecular bone extracted from high-resolution magnetic
  resonance imaging of human trabecular bone: comparison with bone mineral
  density in the prediction of biomechanical strength in vitro,'' {\em Invest
  Radiol}~{\bf 38}(5),  269--80 (2003).

\end{thebibliography}
\bibliographystyle{spiebib}

\end{document}